\documentclass[aps,prb,superscriptaddress,reprint]{revtex4-2}

\usepackage{graphicx}
\usepackage{latexsym}
\usepackage{dcolumn}
\usepackage[utf8]{inputenc}
\usepackage[T1]{fontenc}
\usepackage{tabularx}
\usepackage{hyperref}
\usepackage{color}
\usepackage{placeins}
\usepackage{float}
\usepackage{amsmath}
\usepackage{gensymb}
\usepackage{url}

\begin{document}

\title{Topical Issue ``Dynamics of Systems on the Nanoscale (2021)''. Editorial}

\author{Alexey V. Verkhovtsev}
\email{verkhovtsev@mbnexplorer.com}
\affiliation{MBN Research Center, Altenh\"oferallee 3, 60438 Frankfurt am Main, Germany}
\author{Vincenzo Guidi}
\affiliation{Dipartimento di Fisica e Scienze della Terra, Università degli Studi di Ferrara, Via Saragat 1, 44122 Ferrara, Italy}
\author{Nigel J. Mason}
\affiliation{School of Physics and Astronomy, University of Kent, Canterbury CT2 7NH, United Kingdom}
\author{Andrey V. Solov'yov}
\affiliation{MBN Research Center, Altenh\"oferallee 3, 60438 Frankfurt am Main, Germany}


\begin{abstract}
Exploration of the structure formation and dynamics of animate and inanimate matter on the nanometer scale is a highly interdisciplinary field of rapidly emerging research. It is relevant for various molecular and nanoscale systems of different origins and compositions and concerns numerous phenomena originating from physics, chemistry, biology, and materials science. This topical issue presents a collection of research papers devoted to different aspects of the Dynamics of Systems on the Nanoscale. Some of the contributions discuss specific applications of the research results in several modern and emerging technologies, such as controlled nanofabrication with charged particle beams or the design and practical realization of novel gamma-ray crystal-based light sources. Most works presented in this topical issue were reported at the joint Sixth International Conference ``Dynamics of Systems on the Nanoscale'' and the tenth International Symposium ``Atomic Cluster Collisions'' (DySoN--ISACC 2021), which were held in Santa Margherita Ligure, Italy, in October 2021.
\end{abstract}

\maketitle

\section*{Introduction}
\label{Introduction}

\begin{sloppypar}
Understanding the dynamics of systems on the nanoscale forms the core of a multidisciplinary research area addressing many challenging interdisciplinary problems at the interface of physics, chemistry, biology, and materials science \cite{DySoN_book_2022}.
Common to these interdisciplinary scientific problems is the central role of the structure formation and dynamics of animate and inanimate matter on the
nanometer scale.
\end{sloppypar}

\begin{sloppypar}
Many examples of complex many-body systems of micro- and nanometer-scale size exhibit unique features, properties, and functions.
These systems may have very different nature and origins, e.g. atomic and molecular clusters, biomolecules, ensembles of nanoparticles, nanostructures, nanomaterials,
biomolecular and mesoscopic systems. A detailed understanding of the structure and dynamics of these systems on the nanometer scale is a fundamental and challenging task, the solution of which is required by nano- and biotechnologies, in the development of new materials with unique properties, by the plasma and pharmaceutical industries, and across many disparate fields of medicine.
Detailed understanding and description of these structures, interactions, properties, functions and dynamics are at the core of a new interdisciplinary field of Meso-Bio-Nano (MBN) Science \cite{SolovyovBook2017a} that lies at the intersection of physics, chemistry and biology and establishes new methods for the exploration of the dynamics of systems across various size and time scales.
\end{sloppypar}

The range of open, challenging scientific problems (topical areas) in this research field is very broad. They include:

\begin{sloppypar}
\begin{itemize}
\item Structure and dynamics of molecules, clusters and nanoparticles;
\item Cluster and biomolecular ensembles, composite systems;
\item Clustering, self-organization, phase and morphological transitions on the nanoscale;
\item Nanostructured materials, surfaces and interfaces;
\item Reactivity and nanocatalysis;
\item Radiation-induced chemistry;
\item Irradiation-driven transformations, damage and fabrication of MBN systems;
\item Propagation of particles through media;
\item Biomedical and technological applications of radiation;
\item Emerging technologies: novel light sources, controlled nanofabrication, functionalized materials, etc.
\end{itemize}
\end{sloppypar}

A thorough understanding of MBN systems allows the exploitation of novel phenomena on the nanoscale leading to an optimization of existing processes such as nano\-catalysis \cite{Liu2018,Gelle2020} or the exploration of novel applications involving radiation. The latter encompass controlled nanofabrication technologies such as the focused electron- or ion-beam induced deposition (FEBID or FIBID, respectively) \cite{Utke_book_2012, DeTeresa-book2020}, design and practical realization of novel gamma-ray crystal-based light sources \cite{Channeling_Book2016, Novel_LSs_book_2022, Novel_LSs_review_EPJD_2020}, and novel radiotherapy treatments \cite{IBCT_Book2017}.

\begin{sloppypar}
The premier conference in this field, the international conference entitled ``Dynamics of Systems on the Nanoscale'' (DySoN) \cite{DySoN_conf_website}, started in 2010. Six DySoN conferences have been held so far, and several thematically related topical issues have been published \cite{DySoN2012, DySoN2016, DySoN2018_Editorial}.
\end{sloppypar}

\begin{sloppypar}
The most recent DySoN conference took place in Santa Margherita Ligure, Italy, in October 2021. The conference was organized jointly with the tenth International Symposium ``Atomic Cluster Collisions'' (ISACC) \cite{ISACC_website} under the title ``DySoN--ISACC 2021 Conference''. The conference covered experimental, theoretical and applied aspects of all the aforementioned topics. The utilization of advanced computational techniques and high-performance computing for studying the aforementioned phenomena and effects was also widely discussed.
\end{sloppypar}

The DySoN conference series provides a platform to host discussions about current research, technological challenges and related initiatives within the aforementioned topical areas. The latest DySoN conference highlighted the breakthroughs achieved within the European H2020-MSCA-RISE projects ``RADON'' \cite{RADON_website} and ``N-Light'' \cite{N-Light_website} and the COST Action CA20129 ``Multiscale Irradiation and Chemistry Driven Processes and Related Technologies'' (MultIChem) \cite{MultIChem_website}.

This Topical Issue provides a snapshot of the current research activities in the field of Dynamics of Systems on the Nanoscale, ranging from fundamental research on elementary atomic and molecular mechanisms to studies at a more applied level, covering innovative theoretical, experimental and computational modeling techniques. A particular focus lies on the dynamics of molecules, clusters and nanoparticles; structure and dynamics of biomolecular systems; nanosystems; irradiation-driven transformations involving molecular systems; nanostructured materials; as well as channeling phenomena in oriented crystals and the design of novel light sources. Most contributions to this topical issue have been reported at the DySoN-ISACC 2021 Conference. In this Editorial, we present a brief overview of all the contributions.

\section*{Dynamics of molecules, clusters and nanoparticles}

Several papers in the Topical Issue are devoted to the study of dynamic properties of selected molecular and cluster systems.

\begin{sloppypar}
The photofragmentation dynamics of a van der Waals complex ArBr$_2$ was studied theoretically in Ref.~\cite{GarciaAlfonso_EPJD.76.79}.
The noble gas--halogen van der Waals molecules, where weakly bound rare gas atoms interact with a strong chemical bond like Br--Br, are considered ideal
systems to study molecular energy transfer mechanisms and how molecular properties influence intramolecular dynamics.
In Ref.~\cite{GarciaAlfonso_EPJD.76.79} two different theoretical methods were employed, namely, the trajectory surface hopping (TSH) and the quasi-classical trajectory method (QCTM).
Both methods enabled studies of the dynamical behavior of ArBr$_2$ and the calculation of several observables such as the lifetime, exit channel, rotational energy, and maximum angular momentum of Br$_2$. The results obtained were compared with previous experimental and theoretical studies, while new results were also reported for cases that have not been considered previously.
\end{sloppypar}

\begin{sloppypar}
In Ref.~\cite{PadaShit_EPJD.76.238} molecular dynamics (MD) simulations were performed to explain how the addition of metallic (silver) nanoparticles to an aqueous nanofluid increase its thermal conductivity with temperature. On the basis of the performed MD simulations, the thermal conductivity of a nanofluid in the presence of nanometer-sized spherical silver nanoparticles was calculated using the Green–Kubo framework.
\end{sloppypar}

\section*{Structure and dynamics of biomolecular systems}

Another representative area of DySoN concerns investigations of the structure and dynamics of biomolecular systems. Possible case studies include proteins, DNA, and lipid bilayers, both as isolated systems and in assemblies.
In this Topical Issue, this research area was represented by several theoretical contributions \cite{Hungerland_EPJD.76.154, Schuhmann_EPJD.76.126, Hanic_EPJD.76.198}.

\begin{sloppypar}
The paper \cite{Hungerland_EPJD.76.154} reported results of the atomistic MD-based study of folding$\leftrightarrow$unfolding dynamics of solvated alanine polypeptides.
The atomistic description of the folding process of a structureless chain of amino acids to a functioning protein is still considered challenging for computational biophysics. An in-depth understanding of protein-folding can be achieved through atomistic MD accounting for the solvent effects, combined with the theoretical description of conformational changes in shorter polypeptide chains. The paper \cite{Hungerland_EPJD.76.154} studied the folding transitions in short polypeptide chains made of 11 alanine residues in explicit solvent employing all-atom MD. The multiple observed folding$\leftrightarrow$unfolding events of the peptide were interpreted as a dynamic process and rationalized through the analysis of the potential energy surface of the system. It was demonstrated that the dynamics of alanine polypeptide folding is governed by the backbone dihedral angles and involves the formation of spontaneously folded $\alpha$-helices, which emerge and live for $\sim 1-20$ picoseconds. The helical content within the polypeptide was quantified at different temperatures through a statistical mechanics approach, which showed a reasonable agreement with the results of MD simulations and experiments performed for alanine-rich peptides.
\end{sloppypar}

Another study \cite{Schuhmann_EPJD.76.126} investigated computationally structural and dynamical differences between two enzymes, neutrophil elastase and cathepsin~G. A methodology established in the paper allowed similarities between the protein trajectories describing proteins' temporal evolution to be quantified with accounted for the varying number of amino acid residues comprising each structure. The results reported in Ref.~\cite{Schuhmann_EPJD.76.126} indicated slight differences in the behavior of the active sites of neutrophil elastase and cathepsin~G in the solvent. These subtle changes can result in differences in the general behavior responsible for the different specificity of the two enzymes.

Reference~\cite{Hanic_EPJD.76.198} studied energetic differences of avian cryptochromes 4 proteins. Cryptochromes are a class of light-absorbing proteins that are a part of the circadian rhythm of many animals but play a central role in the magnetosensing of migratory birds.
The study \cite{Hanic_EPJD.76.198} employed classical MD simulations of cryptochrome 4a from five avian species to reveal if any of the cryptochromes feature peculiarities in their internal energetics. The five avian cryptochrome 4a proteins from pigeon, European robin, zebra finch, chicken, and Eurasian blackcap were found to be very similar with respect to their intra-energetic behaviors. Some minor differences between the cryptochromes have been ascribed to the site of specific structural differences. Particular attention was paid to accounting for the interaction of the protein with the solvent.It has been revealed that the solvent can significantly stabilize the chromophore flavin adenine dinucleotide inside the cryptochrome 4a scaffold.

\section*{Irradiation-driven transformations involving mo\-lecular systems}

Irradiation-driven transformations and chemistry induced by the interaction of different types of radiation (X-rays, electrons, ion beams) with molecular systems
is another representative area of DySoN, which is closely linked to many modern and emerging technologies.
For instance, irradiation-driven chemistry (IDC) processes play an important role in ion-beam radiotherapy \cite{Linz2012_IonBeams, schardt2010heavy} that exploits the ability of charged heavy particles to inactivate living cells due to the induction of complex DNA damage \cite{IBCT_Book2017, Surdutovich_2014_EPJD.68.353, Nakano_2022_PNAS}.
IDC transformations in molecular films have been studied in relation to astrochemistry \cite{Mifsud_2021_EPJD.75.182, Mifsud_2022_PCCP.24.10974}. Such transformations occur during the formation of cosmic ices in the interstellar medium due to the interplay of molecular adsorption on a surface and surface irradiation \cite{Tielens_2013_RMP.85.1021}.
Electron-irradiation induced chemistry of organometallic molecules is central to focused electron beam induced deposition (FEBID) -- a technology for the controllable fabrication of complex nanostructures with nanometer resolution \cite{Utke_book_2012, DeTeresa-book2020, Winkler_2019_JAP_review, Huth_2021_JAP_review}.

Several experimental and theoretical papers in this Topical Issue are devoted to electron-induced transformations of molecular systems in the gas and condensed phases.

\begin{sloppypar}
Electron impact-induced fragmentation of free anthracene molecules was studied experimentally in Ref.~\cite{vanderBurgt_EPJD.76.60}. It was shown that double ionization of anthracene molecules by 70~eV electron impact results in a number of prominent fragmentations producing two singly ionized fragments. In the experiment, ionized fragments were detected using a reflectron time-of-flight mass spectrometer. A field-programmable gate array was used for the timing and recording of mass spectra on an event-by-event basis. A detailed model of the coincidence data acquisition was developed, enabling the authors of Ref.~\cite{vanderBurgt_EPJD.76.60} to reliably obtain a map of true coincidences. The measurements showed that fragmentations for which the total number of carbon atoms in the two singly ionized fragments is even are generally significantly stronger than fragmentations for which the total is odd.
\end{sloppypar}

\begin{sloppypar}
The paper \cite{Mifsud_EPJD.76.87} was devoted to the study of radiation chemistry of H$_2$O astrophysical ice analogs. The paper reported the results of an in-depth study of the 2~keV electron irradiation of amorphous solid water (ASW), restrained amorphous ice (RAI) and the cubic (Ic) and hexagonal (Ih) crystalline phases at 20~K to further uncover any potential dependence of the radiation physics and chemistry on the solid phase of the ice. Mid-infrared spectroscopic analysis of the four investigated H$_2$O ice phases revealed that electron irradiation of the RAI, Ic, and Ih phases resulted in their amorphization (with the latter undergoing the process more slowly) while ASW underwent compaction. The abundance of hydrogen peroxide (H$_2$O$_2$) produced as a result of the irradiation was also found to vary between phases, with yields being highest in irradiated ASW. This observation is the cumulative result of several factors, including the increased porosity and quantity of lattice defects in ASW, and its less extensive hydrogen-bonding network. The results of Ref.~\cite{Mifsud_EPJD.76.87} have astrophysical implications, particularly regarding H$_2$O-rich icy interstellar and Solar System bodies exposed to radiation fields and temperature gradients.
\end{sloppypar}

\begin{sloppypar}
A study~\cite{Pintea_EPJD.76.160} focused on the fragmentation of a widely studied FEBID precursor molecule iron pentacarbonyl, Fe(CO)$_5$, during collisions with low-energy electrons. The paper combined experimental mass spectroscopy techniques with \textit{ab initio} calculations using Quantemol-N software package to study electron scattering from Fe(CO)$_5$ and present the fragmentation pathways and channel distributions for each resulting negative ion at low electron energies. The velocity-sliced map imaging (VMI) technique was employed to characterize the fragment anions produced as a result of the dissociative electron attachment (DEA) process. The Quantemol-N package as a standalone was used to study collision processes of low-energy electrons with Fe(CO)$_5$ molecules, including elastic, electronic excitation, and DEA cross sections.
\end{sloppypar}

Investigation of the dynamics of molecular precursors for nanofabrication technologies, such as FEBID, is highly relevant for several modern technological applications, e.g. 3D nano-printing. Molecular interactions, the formation of complex molecular structures, or their aggregation and transformations of these systems under different thermal and irradiation conditions or various external stresses or external fields are exemplary case studies in this research area.
In Ref.~\cite{Prosvetov_EPJD.77.15} the role of thermal effects in the FEBID of Me$_2$Au(tfac) was studied by means of irradiation-driven molecular dynamics simulations using the MBN Explorer \cite{Solovyov2012c} and MBN Studio \cite{Sushko2019} software packages. The FEBID of Me$_2$Au(tfac), a commonly used precursor molecule for the fabrication of gold-containing nanostructures, was simulated at different temperatures in the range of $300-450$~K. The deposit's structure, morphology, growth rate, and elemental composition at different temperatures were analyzed. The fragmentation cross section for Me$_2$Au(tfac) was evaluated on the basis of the cross sections for structurally similar molecules. Different fragmentation channels involving the dissociative ionization (DI) and DEA mechanisms are considered. The conducted simulations of FEBID confirmed the experimental observations that deposits consist of small gold clusters embedded into a carbon-rich organic matrix. The simulation results indicated that accounting for both DEA- and DI-induced fragmentation of all the covalent bonds in Me$_2$Au(tfac) and increasing the amount of energy transferred to the system upon fragmentation increases the concentration of gold in the deposit. The simulations reported in Ref.~\cite{Prosvetov_EPJD.77.15} predicted an increase in Au:C ratio in the deposit from 0.18 to 0.32 upon the temperature increase from 300 to 450~K, being within the range of experimentally reported values.

\section*{Nanostructured materials}

Nanostructured materials are materials with the characteristic size of structural elements of the order of or less than a few hundred nanometers, at least in one dimension. Examples include nanocrystalline materials, nanofibers, nanotubes, and nanoparticle-reinforced nanocomposites. Atomistic modeling of nanosystems utilizes atoms as the elementary building blocks, providing atomic resolution in the computational studies of materials' structure and properties. Such computational investigations, combined with rapid advances in the synthesis of nanostructured materials, have enabled a new active area of materials research that strives to characterize and predict materials with enhanced or unique properties.

In this Topical Issue, nanostructured materials and nanosystems were investigated under several different conditions.

In Ref.~\cite{Dai_EPJD.76.173} the stability of Cd$_8$O$_8$ columnar nanowires and Na-doped Cd$_8$O$_8$ nanowires, along with their electronic properties when under strain, were studied by means of \textit{ab initio} calculations. The results indicated that Cd$_8$O$_8$ nanowires exhibit characteristics of direct gap semiconductors, while the doped nanowires are metallic. It was also shown that both compressive and tensile strains increase the band gap of semiconductor nanowires. The effect of compressive strain on the work function of metal nanowires was found to decrease linearly with an increase in strain, while the work function increases proportionally to the tensile strain. The reported results may have potential applications in optoelectronic devices.

A study \cite{Fan_EPJD.76.212} investigated theoretically optical nonreciprocity in the quantum dot–metal nanoparticle (QD-MNP) hybrid system inside an asymmetric cavity.
Optical nonreciprocity and nonreciprocal devices have attracted significant research interest in recent years due to their important applications in optical communication, information processing, etc. According to the outcomes of Ref.~\cite{Fan_EPJD.76.212} due to the optical nonlinear effect, the input-output relationship of the cavity shows a bistable S-shaped curve, which can be employed to achieve optical nonreciprocity in the asymmetric cavity system. Nonlinear interaction between photons and QD excitons can be strengthened by using the surface plasmon effect in the QD-MNP nanostructure, providing the possibility of achieving a better performance of optical nonreciprocity compared with the single QDs. Not only is the light transmittance enhanced in one direction while further suppressed in the reverse direction, but also the intensity threshold for generating optical nonreciprocity is reduced in the hybrid surface plasmon–exciton cavity system. The results reported in Ref.~\cite{Fan_EPJD.76.212}  may provide references for novel design and experimental realization of nonlinearity-based optical nonreciprocal devices.

The problem of field emission based on a homogeneous array of nanometer-long carbon nanotubes was addressed theoretically in Ref.~\cite{Sadykov_EPJD.77.9}. The length of these nanotubes differs from several nanometers to several hundreds of nanometers. The particle transmission function was obtained considering the difference of potentials on the ends of nanotubes in the range $U = 2 - 3.5$~V. Knowing the particle transmission function, the emission current was calculated.
A linear dependence of current $I$ on the field strength $W$ and a linear dependence of the function $I/W^2$ on the inverse value of field strength $1/W$ have been established. Qualitative matching to experimental results of emission current was obtained when the voltage was kept less than the threshold value of 260~V and the distance between cathode and anode was kept constant and equal to 5~mm.

\section*{Channeling phenomena and the design of novel light sources}

\begin{sloppypar}
Several contributions to this Topical Issue have reported recent advances in theoretical and experimental studies of the channeling phenomenon and radiation produced by ultra-relativistic charged particles in oriented crystals.
This field of research is very promising from the viewpoint of creating new light sources in the sub-Angstrom range of radiation wavelength \cite{Novel_LSs_book_2022, Novel_LSs_review_EPJD_2020}, which may have a broad range of applications.
\end{sloppypar}

\begin{sloppypar}
In Ref.~\cite{Sushko_EPJD.76.166} the feasibility of novel gamma-ray light sources based on the channeling phenomenon of ultra-relativistic charged particles in oriented crystals was demonstrated by means of rigorous numerical modeling that accounts for the interaction of a projectile with all atoms of the crystalline environment. Accurate predictions were provided for the brilliance of radiation emitted in a diamond-based crystalline undulator by a 10 GeV positron beam available at present at the SLAC facility. It was shown that in the photon energy range $\hbar \omega \gtrsim 1$~MeV the brilliance is higher than that predicted in the Gamma Factory proposal for CERN.
\end{sloppypar}

The paper \cite{Sushko_EPJD.76.236} provided an explanation of the key effects behind the deflection of ultra-relativistic electron beams by means of oriented 'quasi-mosaic' bent crystals (qmBC). It was demonstrated that it is essential to characterize the specific geometry of the qmBC and its orientation with respect to a collimated electron beam as well as its size and emittance for an accurate quantitative description of experimental results on the beam deflection by such crystals. In the case study considered in Ref.~\cite{Sushko_EPJD.76.236}, a detailed analysis of a recent experiment at the SLAC facility was presented. The developed methodology has enabled a much better understanding of the peculiarities in the measured distributions of the deflected electrons. This achievement constitutes important progress in the efforts toward the practical realization of novel gamma-ray crystal-based light sources and puts new challenges for the theory and experiment in this research area.

In Ref.~\cite{Backe_EPJD.76.143}, Monte Carlo simulations have been performed for 855~MeV and 6.3~GeV electrons channeling in single silicon crystals along circular bent (111) planes. The aim was to identify the critical experimental parameters which define the volume-deflection and volume-capture characteristics, particularly the angular alignment of the crystal with respect to the nominal beam direction. A continuum potential framework has been utilized. The simulation results were compared with experiments. It was demonstrated that the assumption of an anticlastic bending of the crystal bent on the principle of the quasi-mosaic effect is not required to reproduce the gross features of the experimental observations for the two examined examples.

In another study \cite{Backe_EPJD.76.153}, Monte Carlo simulations have been performed for 855~MeV electrons channeling in (110) planes of a single diamond crystal using the continuum potential framework. Both the transverse potential and the angular distributions of the scattered electrons at screened atoms were based on the Doyle–Turner scattering factors, which were extrapolated with the functional dependence of the Molière representation to large momentum transfers.
The dynamics of the particle in the continuum transverse potential was described classically. Results of the channeling process were presented in terms of instantaneous transition rates as a function of the penetration depth, indicating that channeling can be described by a single exponential function only after about 15 $\mu$m when the equilibration phase has been reached. As a byproduct, improved drift and diffusion coefficients entering the Fokker–Planck equation have been derived.

Bent crystals are a compact and versatile tool to manipulate ultra-relativistic particle beams in accelerators. Indeed, the unrivaled steering power achievable by exploiting the channeling of particles between the atomic planes is comparable to that of a $10^2 - 10^3$ Tesla magnetic dipole. Since the first experiments in the 1980s, extensive research has been delivering important results for new physics experiments and applications in accelerators. The substantial technological development, accompanied by reliable Monte Carlo simulations, has increased the steering efficiency from a few percent to the intrinsic maximum efficiency of ca. 80\%, only being limited by the scattering of the particles with nuclei inside the crystal. The paper by Romagnoni \textit{et al.} \cite{Romagnoni_EPJD.76.135} introduced the two-year (2022–2023) project GALORE of INFN, which aims to assess, experimentally, the possibility of overcoming this hard limit by developing a new generation of bent crystals, featuring an innovative geometry characterized by a crystalline microstructure which influences particle dynamics in the crystal lattice to boost efficiency close to 100\%. The manufacturing process should exploit well-established techniques developed for silicon microelectronics, and experimental testing will be carried out on the beamlines in CERN North Area with high-energy hadron beams. The success of this project can strongly impact the employment of bent crystals in frontier energy accelerators, boosting the performance of already proposed schemes and enabling the use of bent crystals in particle physics experiments in accelerators.

\begin{sloppypar}
A study~\cite{Backe_EPJD.76.150} reported a design study for a positron beam with an energy of 500~MeV to be realized at the applied physics area of the Mainz Microtron MAMI. Positrons will be created after pair conversion of bremsstrahlung, produced by the 855~MeV electron beam at MAMI in a tungsten converter target. Two conceivable geometries, namely (i) pair conversion in the bremsstrahlung converter target itself and (ii) bremsstrahlung pair conversion in a separated lead foil were proposed, the former being presented in some detail.
At an accepted positron bandwidth of 1 MeV, spots are expected vertically with an angular spread of 0.064 mrad and a size of 5.0 mm (FWHM), and horizontally with an angular spread of 0.64 mrad and a size of 7.7 mm (FWHM). The positron yield amounts to 13.1 per second, 1 MeV positron energy bandwidth, and 1 nA electron beam current.
\end{sloppypar}

\section*{Concluding remarks}
To conclude this Editorial, let us mention a crucial issue that concerns the tools needed for a systematic approach to the field of research embraced by MBN Science. Apart from theoretical models and experimental studies, these tools also include computational aspects. MBN Science, as a research field, has only emerged recently from, primarily, atomic, molecular, and cluster physics, together with the development of powerful computers and advanced computational techniques. The computational aspect of MBN Science provides the methodology for revealing novel features of structure and dynamics of nano\-scopic and mesoscopic mo\-lecular systems. It also requires a high level of interdisciplinary collaboration since similar computational methodologies can be easily adapted to mo\-lecular systems of very different nature and origins.

To fully understand and exploit all the richness and complexity of the MBN world, especially its all-atom dynamics, one needs to engage many disciplines ranging from physics and chemistry to materials and life science, exploiting technologies from software engineering and high-performance computing. This challenge led to the formulation and development of the multiscale software packages {MBN Explorer} \cite{SolovyovBook2017a,Solovyov2012c} and MBN Studio \cite{Sushko2019}, designed to become powerful and universal instruments of computational research in the field of MBN Science.
Many contributions in this Topical Issue \cite{Hungerland_EPJD.76.154, Schuhmann_EPJD.76.126, Hanic_EPJD.76.198, Prosvetov_EPJD.77.15, Sushko_EPJD.76.166, Sushko_EPJD.76.236}
highlighted case studies conducted with these packages and demonstrated their application to various topical areas in MBN Science.

\begin{acknowledgments}
The authors acknowledge partial financial support by the European Commission through the RADON (GA 872494) and  N-LIGHT (GA 872196) projects within the H2020-MSCA-RISE-2019 call, and the Horizon Europe EIC Pathfinder Project TECHNO-CLS (Project No. 101046458). This article is also based upon work from the COST Action CA20129 MultIChem, supported by COST (European Cooperation in Science and Technology).
\end{acknowledgments}

\end{document}